\documentclass[preprint,amsmath,amssymb,superscriptaddress]{revtex4-1}

\usepackage{graphicx}
\usepackage{dcolumn}
\usepackage{bm}
\usepackage{amsfonts} 
\usepackage{color}

\begin{document}
\title{Scalable quantum computer with superconducting circuits in the ultrastrong coupling regime}
\author{Roberto Stassi} \email{roberto.stassi@riken.jp}
\affiliation{Theoretical Quantum Physics Laboratory, RIKEN Cluster for Pioneering Research, Wako-shi, Saitama 351-0198, Japan}
\affiliation{Dipartimento di Scienze Matematiche e Informatiche,
Scienze Fisiche e Scienze della Terra, Universit\`a di Messina, I-98166 Messina, Italy}
\author{Mauro Cirio}
\affiliation{Graduate School of China Academy of Engineering Physics,\\ Haidian District, Beijing, 100193, China}
\author{Franco Nori}
\affiliation{Theoretical Quantum Physics Laboratory, RIKEN Cluster for Pioneering Research, Wako-shi, Saitama 351-0198, Japan}
\affiliation{Physics Department, The University of Michigan,\\ Ann Arbor, Michigan 48109-1040, USA.}%

\date{\today}

\begin{abstract}

So far, superconducting quantum computers have certain constraints on qubit connectivity, such as nearest-neighbor couplings.
To overcome this limitation, we propose a scalable architecture to \textit {simultaneously} connect several pairs of \textit{distant} qubits via a dispersively coupled quantum bus. The building-block of the bus is composed of orthogonal coplanar waveguide resonators connected through ancillary flux qubits working in the ultrastrong coupling regime. This regime activates virtual processes that boost the effective qubit-qubit interaction, which results in quantum gates on the nanosecond timescale. The interaction is switchable and preserves the coherence of the qubits.

\end{abstract}

\maketitle

\section*{Introduction}
Superconducting circuits are a very promising hardware platform for quantum computers with capabilities beyond the ones of classical computers (see, e.g., \cite{Barends:2014fu, Mohseni:2017jw, IBMquantum, Boixo:2018im, arute2019quantum} and references therein). A basic requirement to perform quantum logic gates is to have controllable interactions among qubits (e.g., \cite{grajcar2006, LiuY2006, Plantenberg:2007bs, ashhab2008int}). Obviously, quantum computers benefit from higher and better connectivity among qubits, and this becomes more challenging to achieve as the system is scaled up. 
Unfortunately, so far, superconducting quantum computers have certain constraints on qubit connectivity, such as nearest-neighbor couplings \cite{Linke:2017jj}. Although the distant interaction between two or more qubits, mediated by a cavity bus, has been demonstrated (e.g., \cite{DiCarlo:2009jaa, DiCarlo:2010js,song2019generation}), this scheme is not convenient to connect \textit{many} pairs of distant qubits \textit{simultaneously} in a superconducting quantum computer. Indeed, in this case, the qubit-qubit interaction is activated by tuning qubit frequencies, leading to possible unwanted couplings and to a reduction of the coherence time of the qubits.\\
In addition to applications to quantum computing, superconducting circuits are a very versatile platform to investigate new quantum phenomena and to engineer quantum devices (e.g., \cite{You:2011jj, Devoret1169, Martinis:2015ig, Wendin:2017bw, preskill2018, Brecht:2016km, Stassi:2018bk, gu2017microwave}). Note that the coupling between a superconducting artificial atom (e.g., \cite{YouJ2007, Steffen2010, Yan:2016df, Corcoles2011}) and a resonator can be a significant fraction of the atom  and cavity bare energies (e.g., \cite{Bourassa:2009gy,Niemczyk:2010gv,Yoshihara:2017bia,FornDiaz:2016bo,Chen:2017dn,di2019resolution}). In this ultrastrong coupling regime, the usual Jaynes-Cummings approximation breaks down and the counter-rotanting terms must be taken into account \cite{Kockum:2019ky, FornDiaz2018}.\\ 
Here, we theoretically propose a \textit{scalable architecture to simultaneously couple pairs of distant superconducting qubits}. The building block of this architecture is composed of three waveguides. Two of them (${\rm C}_1$ and ${\rm C}_2$), see Fig.\,1a, are directly connected to the qubits (${\rm q_a}$ and ${\rm q_b}$) that form the computational basis, while a third (${\rm C}_3$) is connected to the first two in a $\Pi$-shape form. At the intersection point, the interaction is mediated by ancillary flux qubits (${\rm f}_1$ and ${\rm f}_2$) in the ultrastrong coupling regime. All components of the $\Pi$-connector are on resonance with each other. However, the two qubits, ${\rm q_a}$ and ${\rm q_b}$, are detuned with respect to the eigenenergies of the bus. The last condition guarantees that the coupling between qubits is mediated by virtual excitations, thereby not affecting their coherence. Moreover, the bus takes advantage of the counter-rotating terms activated by the ultrastrong coupling, enhancing the coupling between the qubits. This allows to perform fast two-qubit gates on nanosecond timescales. To achieve scalability, these building blocks can be arranged in an array, see Fig.\,1b, so that every qubit is connected with each other. Couplings between qubits can be switched on and off by tuning the ancillary flux qubit frequencies on and off resonance with the waveguides. Importantly, this allows the qubits ${\rm q_a}$ and ${\rm q_b}$ to remain in their optimal working point, preserving their coherence times.
\begin{figure}[h]
  \includegraphics[scale=0.33 ]{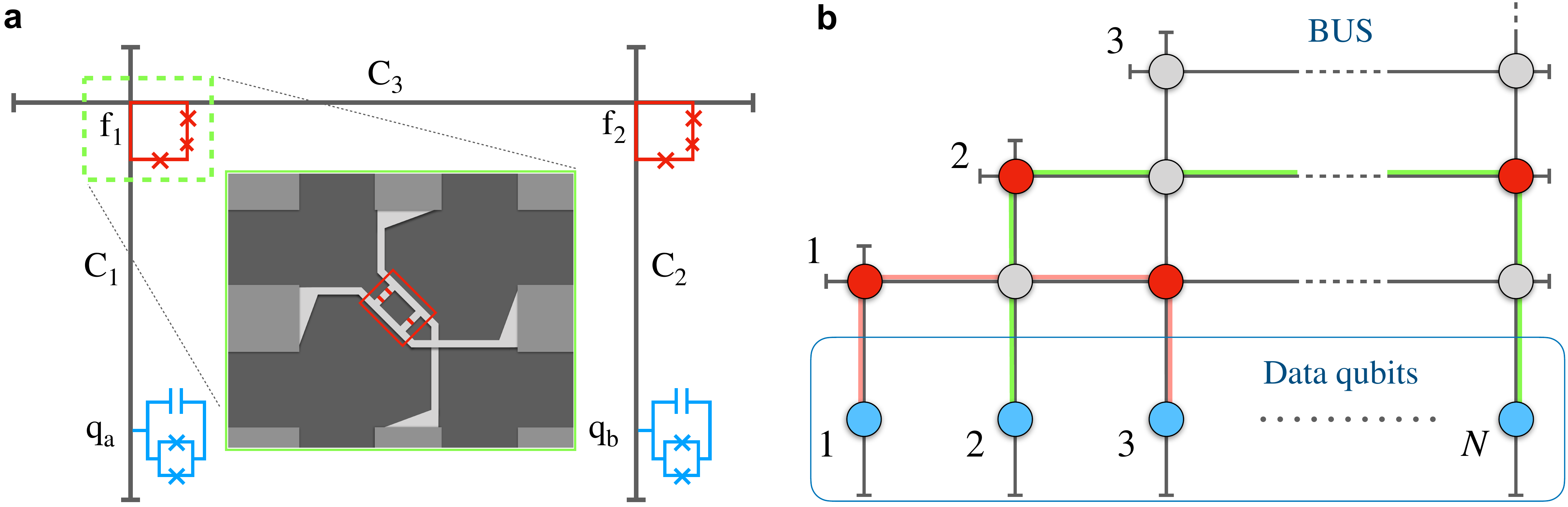}
  \caption{\textbf{$\mathbf\Pi$-connector and scalable architecture.} \textbf{a} Sketch of the $\Pi$-connector. Dark grey lines represent the coplanar waveguide resonators, $\rm{C}_1$, $\rm{C}_2$ and $\rm{C}_3$. Red lines represent the flux qubits, $\rm{f}_1$ and $\rm{f}_2$, connecting the waveguides. Blue lines represent the data qubits (transmon here, but could also be other types), $\rm{q}_a$ and $\rm{q}_b$. The inset inside the green dashed square represents the connection between the flux qubit (red box) and the constriction of the center conductor of the two orthogonal waveguides (light grey). \textbf{b} An array of data qubits (blue disks) at the bottom part are connected through a net of waveguides. At each node, a flux qubit tuned with the waveguides (red disk) mediates the interaction between data qubits.  The grey disks (connector is ``OFF'') denote detuned flux qubits.}
\end{figure}

\section*{Results}
\begin{figure}[h]
  \includegraphics[scale=0.4]{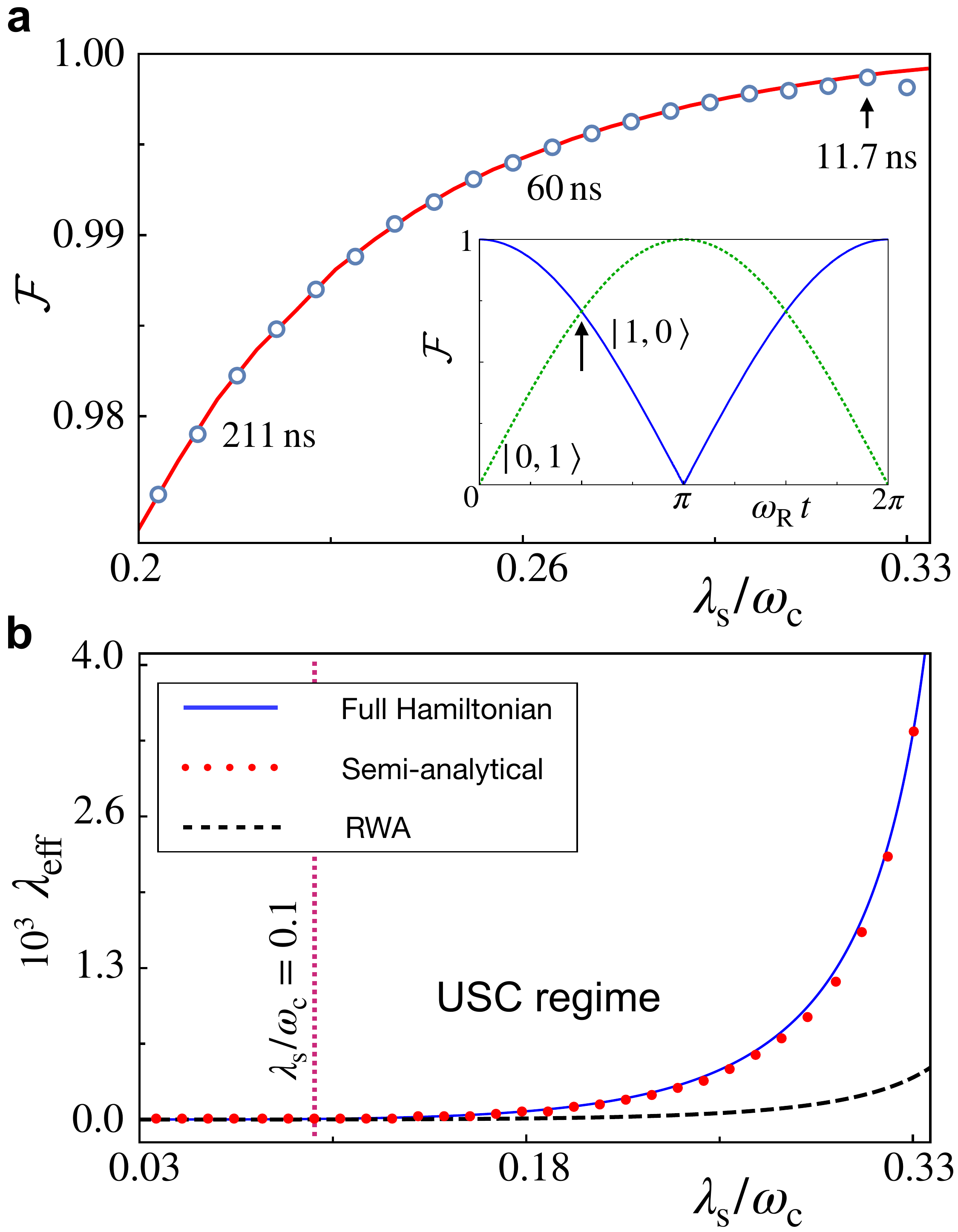}
  \caption{\textbf{Fidelity and effective coupling.} \textbf{a} Average gate fidelity of the $\sqrt{i{\rm SWAP}}$ gate generated by the quantum bus (blue circles) and by a direct qubit-qubit coupling (red solid curve) with coupling strength $\lambda_{\rm eff}$. The chosen frequency transitions of the data qubits is $\omega_{\rm q}=4\,\rm{GHz}$, the relaxation time is $70\,\mu \rm s$ and the pure dephasing time is $92\,\mu \rm s$. Flux qubits have a relaxation time of $20\, \mu \rm s$ and pure dephasing time of $10\, \mu \rm s$; resonators have a $Q$-factor of $5\times 10^5$. The inset shows the fidelity of the $\vert 1,0\rangle$ (blue solid curve) and $\vert 0,1\rangle$ states (green dotted curve). Here the evolution is numerically calculated using the quantum bus, no dissipation is considered. \textbf{b} Effective coupling calculated numerically using the full Hamiltonian $\hat H$ (solid blue curve), dropping the counter-rotating terms (dashed black curve) and calculated using the semi-analytical expression in Eq.\,(\ref{equ2}) (red dots).}
 \end{figure}
The Hamiltonian describing the $\Pi$-connector in Fig.\,1a is $\hat H=\hat H_{\rm qb}+\hat H_{\Pi}+\hat H_{\rm int}$, where $\hat H_{\rm qb}=\frac{1}{2}\sum_{i={\rm a,b}}\omega_{{\rm q}_{i}}\,\hat\sigma_z^{(i)}$ represents both qubit ${\rm q_a}$ and ${\rm q_b}$ $(\hbar=1)$,
\begin{equation}
  \hat H_{\Pi}=\frac{1}{2}\sum_{i=1}^2\omega_{{\rm f}_i}\hat\sigma_z^{(i)}+\sum_{i=1}^3\omega_{\rm c}\hat a^{\dagger}_{(i)}\hat a^{ }_{(i)}+\sum_{i=1}^2\lambda_{\rm s}\,\hat\sigma_x^{(i)}\left(\hat X_i+\hat X_3\right)
  \label{equ1}
\end{equation}
is the Hamiltonian of the ultrastrongly coupled quantum bus, and $\hat H_{\rm int}=\lambda\left(\,\hat\sigma_x^{({\rm a})}\hat X_1+\hat\sigma_x^{({\rm b})}\hat X_2\right)$ represents the interaction between the qubits and the quantum bus. Here, $\hat\sigma_z^{(i)}$ and $\hat\sigma_x^{(i)}$ are Pauli operators for the qubits  ${\rm q_a}$ and ${\rm q_b}$ ($i=a,b$) and for the flux qubits ($i =1,2$), with transition energies $\omega_{{\rm q}_i}=\omega_{\rm q}$ and $\omega_{{\rm f}_i}=\omega_{\rm c}$, respectively. We set the fundamental frequency of all resonators ${\rm C}_k$ to be $\omega_{\rm c}=3\,\omega_{\rm q}$, and we denote the annihilation, creation, and quadrature operators by $\hat a^{ }_{(k)}$, $\hat a_{(k)}^\dagger$, and $\hat X_k=\hat a^{ }_{(k)}+\hat a_{(k)}^\dagger$ ($k=1,2,3$).  Flux qubit $\rm {f_1}$ ($\rm {f_2}$) is ultrastrongly coupled to cavity $\rm C_1$ ($\rm C_2$) and $\rm C_3$, with coupling strength $\lambda_{\rm s}>0.1\,\omega_{\rm c}$. Instead, the coupling strength between qubit $\rm q_{a}$ ($\rm q_{b}$) and resonator $\rm C_1$ ($\rm C_2$) is set to $\lambda=0.05\,\omega_{\rm q}$.  The latter interaction, operating in the dispersive regime, causes a shift in the qubit frequency, 
\begin{equation}
  \tilde{\omega}_{{\rm q}_i}=\omega_{{\rm q}_i}-\lambda^2/(\omega_{\rm c}+\omega_{{\rm q}_i})\,,
\end{equation}
and a dressing in the qubit states \cite{beaudoin2011, Rigetti:2012en}. These qubit dressed states $\{\vert 0\rangle,\vert 1\rangle\}$ are the ones forming our computational basis. We call ``data qubits'' the dressed qubits which are generating the computational basis.\\
The quantum bus provides an effective \textit{XX} interaction between data qubits mediated by virtual excitations, as it is guaranteed by the detuning condition $\omega_c-\omega_{{\rm q}}=2\,\omega_{{\rm q}}$. This interaction causes a two-qubit oscillation between states with one excitation: $\vert 1,0\rangle$ and $\vert 0,1\rangle$. The inset of Fig.\,2a shows the swap from the state $\vert 1,0\rangle$ to the state $\vert 0,1\rangle$ in a time $t=\pi/\omega_{\rm R}$, where $\omega_{\rm R}=2\lambda_{\rm eff}$, and $\lambda_{\rm eff}$ is the effective coupling strength between data qubits. At $t=\pi/2\omega_{\rm R}$, as indicated by the arrow, a maximally entangled Bell state is generated, and a universal $\sqrt{i{\rm SWAP}}$ gate is obtained.  Figure 2a shows the average gate fidelity \cite{poyatos1997,emerson2005}
\begin{equation}
  \mathcal{F}=\int d\Psi\langle\Psi\vert \hat U^{\dagger}_{\sqrt{i{\rm SWAP}}}\,\hat\rho_{\vert\Psi\rangle}\hat U_{\sqrt{i{\rm SWAP}}}\vert\Psi\rangle
  \label{Fid}
\end{equation}
generated by the quantum bus as a function of the ultrastrong coupling $\lambda_{\rm s}$ (blue circles), taking in consideration decoherence channels originating from the components of the bus. Here, $\hat\rho_{\vert\Psi\rangle}$ is the resulting density matrix after evolving the system for a time $t_{\sqrt{i{\rm SWAP}}}=\pi/2\omega_{\rm R}$, under the action of the full Hamiltonian in Eq.\,(\ref{equ1}). The integral in Eq.\,(\ref{Fid}) uses the unitarily invariant measure $d\Psi$ on the state space, normalized such that $\int d\Psi=1$, the operator $\hat U_{\sqrt{i{\rm SWAP}}}$ is the ideal $\sqrt{i{\rm SWAP}}$ gate, and $\vert\Psi\rangle$ is the input state. For the data qubits we choose $T_1=70\,\mu \rm s$ and pure dephasing time $T_{\varphi}=92\,\mu \rm s$ (data are taken from Ref.\,\cite{Rigetti:2012en}, that fulfil our parameter conditions).\\
When the interaction is activated, the ultrastrong coupling induces a small energy-shift of the data qubit, resulting in a \textit{z}-axis single-qubit rotation. This can be compensated using standard procedures \cite{krantz2019}. In our simulations it is compensated by a rotation in the opposite direction. Figure 2a also shows the average fidelity of the $\sqrt{i{\rm SWAP}}$ gate (red solid curve) when the data qubits are directly coupled through the ideal interaction Hamiltonian $\lambda_{\rm eff}\,\hat\sigma_x^{(a)}\hat\sigma_x^{(b)}$. A comparison between the two curves proves that the bus does not affect the coherence of the data qubits, and shows that the only limitation to performance is the intrinsic decoherence of the data qubits. Considering data qubits with transition frequency of $4\,\rm{GHz}$ \cite{Rigetti:2012en}, for $\lambda_{\rm s}=0.32\,\omega_{\rm c}$ the fidelity is $99.87\%$ and the gate time is $11.7\,\rm ns$. Beyond this coupling point, the gate performance degrades due to the hybridization between the computational and bus states. All the dynamics are calculated using the master equation developed in Ref.\,\cite{Stassi:2018bk}, for $T=0$.\\
Every type of superconducting qubit can be used as data qubit in our protocol. Currently, due to their long coherence time \cite{Rigetti:2012en}, transmon qubits are commonly used. However, the low anharmonicity of these artificial atoms could lead to non-negligible detrimental effects on the gate performance. To estimate these effects, we calculated the average gate fidelity using as data-qubit the two lowest states of a three-level system with low anharmonicity. We set the transition frequency between the first and second excited states to $0.8\,\omega_{\rm q}$, where $\omega_{\rm q}$ is now the transition frequency between the ground and the first excited state. All other parameters are as above. In this setting, the average gate fidelity is $99.72\%$ for $\lambda_{\rm s}/\omega_{\rm c}=0.3$ (instead of $99.75\%$ calculated using the two-level system). Moreover, in the absence of decoherence channels, the fidelity is $99.99\%$. These results indicate that the low anharmonicity has a negligible impact on the performance of the gate.\\

\begin{figure}[h]
  \includegraphics[scale=0.4]{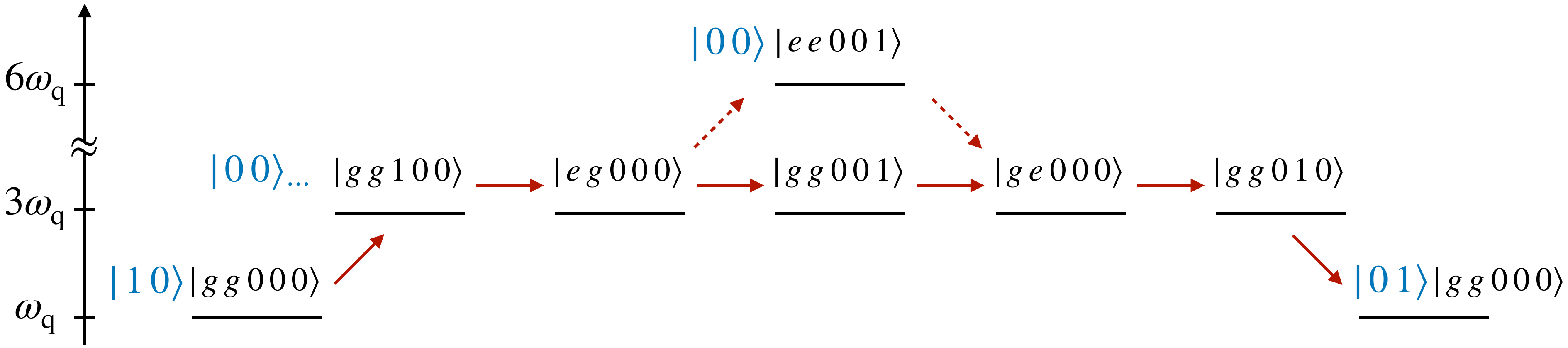}
  \caption{\textbf{Virtual states connecting $\mathbf{q_a}$ and $\mathbf{q_b}$} The main path (solid arrows) connecting the states $\vert 1, 0\rangle$ and $\vert 0, 1\rangle$ (blue states) through the virtual bus states (black states). The order of the labels in the black kets is $\vert {\rm f_1,f_2,C_1,C_2,C_3} \rangle$. The dashed arrows indicate a path due to the counter-rotating terms.
}\label{fig4}
\end{figure}
\textbf{Effective coupling}\\
As explained in the Methods section, to calculate the effective qubit-qubit coupling we perform a projection of the full Hamiltonian $\hat H$ into the ground state of the bus Hamiltonian $\hat H_{\Pi}$. Considering the dispersive regime between data qubits and the bus, neglecting the dressing, the effective coupling becomes 
\begin{equation}
  \lambda_{\rm eff}=\sum_k\frac{g_k^{\rm (1)}g_k^{\rm (2)}}{\omega_{\rm q}-\Delta E_k}\,,
  \label{equ2}
\end{equation}
where $E_k$ and $\vert \tilde k\rangle$ are the eigenenergies and eigenstates of $H_{\Pi}$, and where $g_k^{ (i)}=\lambda\,\langle \tilde k\vert \hat X_i\vert \tilde 0\rangle$ $(i=1,2)$, and $\Delta E_k=E_k-E_0$ \cite{Kyaw2017}. 
 In Fig.\,2b, we numerically computed the effective coupling as a function of $\lambda_{\rm s}$ using the full Hamiltonian $\hat H$, and compared it with Eq.\,(\ref{equ2}). The agreement is very good  in the coupling range under investigation. According to perturbation theory to sixth-order \cite{Garziano:2016jy,Stassi:2017hd,kockum2017}, the virtual processes that provide the main contribution to the qubit-qubit effective interaction, neglecting the dressing, are the ones that connect the state $\vert 1,0\rangle\vert 0\rangle_{\rm b}$ to $\vert 0,1\rangle\vert 0\rangle_{\rm b}$ (where $\vert 0\rangle_{\rm b}=\vert g,g,0,0,0\rangle$) through states with the lowest-energy differences with the initial state, $\vert 1,0\rangle\vert 0\rangle_{\rm b}$. It appears clear now that the main process, Fig.\,3 (red solid arrows), is the one that transfers one excitation through all the elements that compose the bus. In the same diagram, it is also shown a virtual process (red dashed arrows) involving the simultaneous excitation of the flux qubit $\rm f_1$ and the resonator $\rm C_3$, which is activated by the counter-rotating terms in the interacting part of the bus Hamiltonian $\hat H_{\Pi}$.\\
 In the ultrastrong coupling regime, the counter-rotating terms become relevant and activate virtual processes that strongly boost the effective coupling. To prove this, we have numerically calculated the effective coupling after dropping the counter-rotating terms in $\hat H$ (see Fig.\,2b, dashed curve). Comparing this with the results from the full Hamiltonian (blue solid curve), we notice that $\lambda_{\rm eff}(\lambda_{\rm s})$, calculated with the counter-rotating terms, increases much faster compared to the one calculated without it, as a function of the coupling $\lambda_{\rm s}$.  It is standard procedure in perturbation theory to use virtual excitations to derive an effective interaction. These are the virtual photons we are referring to, not the ones in the ultrastrong coupling of cavity QED.\\

\textbf{Switch-on and -off of the effective interaction}\\ 
To realize a properly scalable system, it is important to be able to switch-on and -off the interaction between arbitrary data qubits. We achieve this by tuning (switch-on) and detuning (switch-off) to the bus the transition frequency of the ancillary flux qubit by varying the external flux $\Phi_{\rm ext}=f\,\Phi_0$ threading it \cite{gu2017microwave}. We set the switch-on condition at the optimal bias point, $f\to f_{\rm on}=0.5$, where the flux qubit has a symmetric potential energy and maximum dipole moment ${\rm M}_{\rm on}$ \cite{Liu:2005eo}. To switch-off the interaction we move the flux qubit away from its optimal point, by changing the external flux, $f\to f_{\rm off}$.\\
If we detune $\rm f_1$ and $\rm f_2$ from the $\Pi$-connector in Fig.\,1a, using $f_{\rm off}=0.522$, the flux qubit transition-frequency becomes $\approx 14\,\omega_{\rm f_1}$, and the dipole moment becomes ${\rm M}_{\rm off}=6\times 10^{-2}\,{\rm M}_{\rm on}$ (other parameters are provided in Methods). For $\lambda_{\rm s}=0.3\,\omega_{\rm c}$, the residual interaction is $\lambda_{\rm eff}^{\rm (off)}\approx 2\times\,10^{-11}\omega_{\rm q}$ and the on/off coupling ratio between data qubits is $\approx 6\times 10^{7}$, which is almost independent of $\lambda_{\rm s}$. When the flux qubit $\rm f_2$ is detuned and $\rm f_1$ is tuned, the on/off coupling ratio is $\approx 9\times 10^{3}$. In this case, if the system consists of only two data qubits, no interaction occurs. This happens as the ultrastrong coulpling shifts the frequency of data qubit $\rm q_a$ by a quantity larger than the residual effective coupling. For instance, if $\rm f_{2}$ is detuned and $\lambda_{\rm s}=0.3\,\omega_{\rm c}$, the qubit $\rm q_{b}$ interacts with qubit $\rm q_{a}$ at $\omega_{\rm q_{b}}=\omega_{\rm q}-9.3\times 10^{-4}\,\omega_{\rm q}$, with a residual effective coupling of $\tilde\lambda_{\rm eff}^{\rm (off)}=1.5\times 10^{-7}\omega_{\rm q}$. Note that when the flux qubits are not in the optimal bias point, a charge interaction with the second quadrature of the resonators is activated, but its contribution is negligible \cite{Bourassa:2009gy}.\\

\textbf{Scalable architecture}\\
Figure 1b shows a possible scalable architecture for quantum computation using the $\Pi$-connector. In the bottom part of Fig.\,1b, we represent an array of data qubits  (blue disks). In the upper part we present the quantum bus. At each node, ancillary flux qubits can either couple (red disks) or decouple (grey disks) to the waveguides, depending on their frequency. In this way, it is possible to control the connectivity among arbitrary pairs of data qubits. For example, in Fig.\,1b qubit 1 is connected to qubit 3, and qubit 2 is connected to qubit $N$. It is also possible to connect more than two qubits simultaneously.  If the fundamental mode of the superconducting coplanar resonator is $12\, \rm{GHz}$ \cite{Rigetti:2012en}, and if the distance between two consecutive flux qubits in the resonator is $0.1\,{\rm mm}$ (which could be even shorter), it could be possible to connect about 100 data qubits. The effective interaction among $N$ data qubits in the scalable architecture is described by
\begin{equation}
  \hat H_{\rm I}=\sum_{k=1}^{N-1}\sum_{l=\,k+1}^N k\lambda_{\rm eff}^{kl}\,\hat\sigma_x^{(l)}\hat\sigma_x^{(k)}\,.
  \label{eq3}
\end{equation}
This considers that, i.e., qubit 1 is connected (on or off) with all the other qubits using 1 path, qubit 2 is connected with the rest of the qubits using 2 paths for each qubit, qubit 3 is connected with the remaining qubits using 3 paths, and so on.\\
To evaluate cross-talk, we calculate the interaction of qubit $N$, that is the one with more connections, with all the other data qubits in the off coupling condition, $\lambda_{\rm eff}^{kl}=\lambda_{\rm eff}^{\rm (off)} $. From Eq.\,(\ref{eq3}), we find
\begin{equation}
  \hat H^{(N)}_{\rm I}=\hat\sigma_x^{(N)}\left[\left(N-1 \right)\lambda_{\rm eff}^{\rm (off)}\hat\sigma_x^{(N-1)}+\left(N-2 \right)\lambda_{\rm eff}^{\rm (off)}\hat\sigma_x^{(N-2)}+\ldots+\lambda_{\rm eff}^{\rm (off)}\hat\sigma_x^{(1)}\right]\,.
\end{equation}
Considering the $(N-1)$ data qubits as a single effective qubit which is interacting with qubit $N$, $\hat\sigma_x ^{(k)}\rightarrow\hat\sigma_x$ for all $k\neq N$, we obtain
\begin{equation}
  \hat H^{(N)}_{\rm I}=\frac{N(N-1)}{2}\lambda_{\rm eff}^{\rm (off)}\hat\sigma_x^{(N)}\hat\sigma_x\,.
\end{equation}
Now,
\begin{equation}
  \Lambda_{\rm eff}^{\rm (off)}=\lambda_{\rm eff}^{\rm (off)}N(N-1)/2
\end{equation}

 is  the residual interaction that affects qubit $N$ in the off coupling condition. For $\lambda_{\rm s}=0.3\,\omega_{\rm c}$, using $N=100$ data qubits, the coupler has a numerically measured on/off ratio of $\approx 12,000$.\\
In this architecture, when each pair of data qubits is interacting, all the data qubits have the same frequency-shift and the residual coupling $\tilde\lambda_{\rm eff}^{\rm (off)}$ is active. It is possible to cancel out this small interaction by detuning every pair of interacting data qubits from all other pairs by a quantity larger than the residual interaction. This can be achieved by changing the flux qubit frequency, that in turn changes the data qubit dressing.
 
\section*{Discussion}
By taking advantage of the large coupling between flux qubits and the modes of waveguides or $LC$ resonators, we proposed a scalable architecture which allows to control the coupling between many distant qubits. We numerically showed that the effective coupling is boosted by the counter-rotating terms of the Rabi Hamiltonian, whose contribution become more relevant in the ultrastrong coupling regime. The switch-on and -off of the interaction between data qubits is controlled by the magnetic fluxes threading the flux qubits, which tune their transition frequencies to the bus. Note that the resonant condition among the waveguides and flux qubits in the bus does not have to be perfect, because there is considerable tolerance. In fact, the strength of the effective coupling does not depend on this condition, but it depends on the detuning between each element of the bus and the data qubits, and on the couplings between elements of the bus. However, the resonant condition between data qubits must be satisfied. Unfortunately, current fabrication tolerances do not allow to set the frequency of the qubits precisely, and SQUID loops must be used to tune the data qubit frequencies. However, near-future improvements in the fabrication quality of qubits will eventually allow to take full advantage of this proposal. 
We believe that this architecture might lead to a new generation of quantum computer architectures controlled by elements largely detuned from the data ones, allowing to increase the complexity of the system without affecting the coherence times. A natural evolution could be the connection of a matrix of data qubits through waveguides in a 3D circuit \cite{Brecht:2016km}.

\section*{Methods}
\textbf{Effective coupling}\\ In this section, we derive an effective model to describe the dynamics of two data qubits in contact with a quantum bus. 
We do this by projecting the full dynamics (which takes place in the total Hilbert space $\mathcal{H}$ of both data qubits and bus) into the subspace $\mathcal{H}_\text{eff}=P\mathcal{H}$, where the bus is in the  ground state. Here, $\hat P=\mathbb{\hat I}_{\text{qb}}\otimes\vert \tilde{0} \rangle\langle \tilde{0} \vert$ denotes the projector into the ground state $\vert \tilde{0} \rangle$ of the bus ($\mathbb{\hat I}_{\text{qb}}$ being the identity operator on the data qubits). \\
As a first step, we decompose the total Hamiltonian $\hat{H}$ into a ``diagonal'' contribution $\hat{H}_0$ (which preserves $\mathcal{H}_\text{eff}$, i.e., $[\hat{H}_0,\hat P]=0$) and an ``off-diagonal'' contribution $\hat{V}$ (for which $[\hat{V},\hat P]\neq 0$). By defining a complementary projector $\hat Q$, such that $\hat P+\hat Q=\mathbb{\hat I}$, we can write
\begin{equation}
\begin{array}{lll}
\hat{H}&=&(\hat P+\hat Q)\hat{H}(\hat P+\hat Q)\\
&=&\hat{H}_0 + \hat{V}
\end{array}
\end{equation}
 where $\hat{H}_0=\hat P\hat{H}\hat P+\hat Q\hat{H}\hat Q$ and $\hat{V}=\hat P\hat{H}_{\text{int}}\hat Q+\hat Q\hat{H}_{\text{int}}\hat P$.  The potential $\hat{V}$ can be explicitly written as
\begin{equation}
\begin{array}{lll}
\hat V&=&\displaystyle\sum_k \left[g_k^{(1)}\hat{\sigma}_x^{(a)}\left(
\vert \tilde{k} \rangle\langle \tilde{0} \vert+\vert \tilde{0} \rangle\langle \tilde{k} \vert\right)+ g_k^{(2)}\hat{\sigma}_x^{(b)}\left(\vert \tilde{k} \rangle\langle \tilde{0} \vert+\vert \tilde{0} \rangle\langle \tilde{k} \vert\right)\right]\\
&\simeq&\displaystyle\sum_k \left[g_k^{(1)}\hat{\sigma}_-^{(a)}\vert \tilde{k} \rangle\langle \tilde{0} \vert+g_k^{(2)}\hat{\sigma}_-^{(b)}\vert \tilde{k} \rangle\langle \tilde{0} \vert\right]+ \rm H.c. \,,\\
\end{array}
\end{equation}
where we made a rotating-wave approximation under the assumption that $\vert g_k^{(1)}\vert, \vert g_k^{(2)}\vert\ll\omega_\text{q}, \Delta E_k$.
We further assume to be in a dispersive regime where the detuning between the splitting of the data qubits $\omega_\text{q}$ and the transition energies of the bus $\Delta E_k$ are much larger than the couplings $g_k^{(1)}$ and $g_k^{(2)}$ (i.e., $\vert\omega_q-\Delta E_k\vert\gg \vert g^{(1)}_k\vert, \vert g^{(2)}_k\vert$). In this limit, it is possible to perturbatively define a rotating frame where the dynamics is effectively constrained in $\mathcal{H}_\text{eff}$ (Schrieffer-Wolff transformation). Specifically, a change of frame $\exp[{\hat{S}}]$ (for an anti-Hermitian operator $\hat{S}$ such that $[\hat{H}_0,\hat{S}]=\hat{V}$) allows to define the effective Hamiltonian
\begin{equation}
\label{eq:temp}
\begin{array}{lll}
\hat{H}_\text{eff}=\displaystyle \hat P e^{\hat{S}} \hat{H} e^{-\hat{S}}\hat P
&\simeq&\displaystyle \hat P\hat{H_0}\hat P + \frac{1}{2}\hat P[\hat S,\hat{V}]\hat P\,,
\end{array}
\end{equation}
 at the lowest non-trivial order in $\hat S$. Specifically, by choosing
\begin{equation}
\hat S=\sum_{k>0}\left(\frac{g_k^{(1)}}{\omega_q-\Delta E_k}\hat\sigma^{(a)}_++\frac{g_k^{(2)}}{\omega_q-\Delta E_k}\hat\sigma^{(b)}_+\right)\vert \tilde{0} \rangle\langle \tilde{k} \vert-\text{H.c.}\,,\\
\end{equation}
and computing the commutator $[\hat S,\hat{V}]$ in Eq.~(\ref{eq:temp}), we obtain the effective coupling between the data qubits described in the main text.\\

\begin{figure}[hbt]
  \includegraphics[scale=0.4]{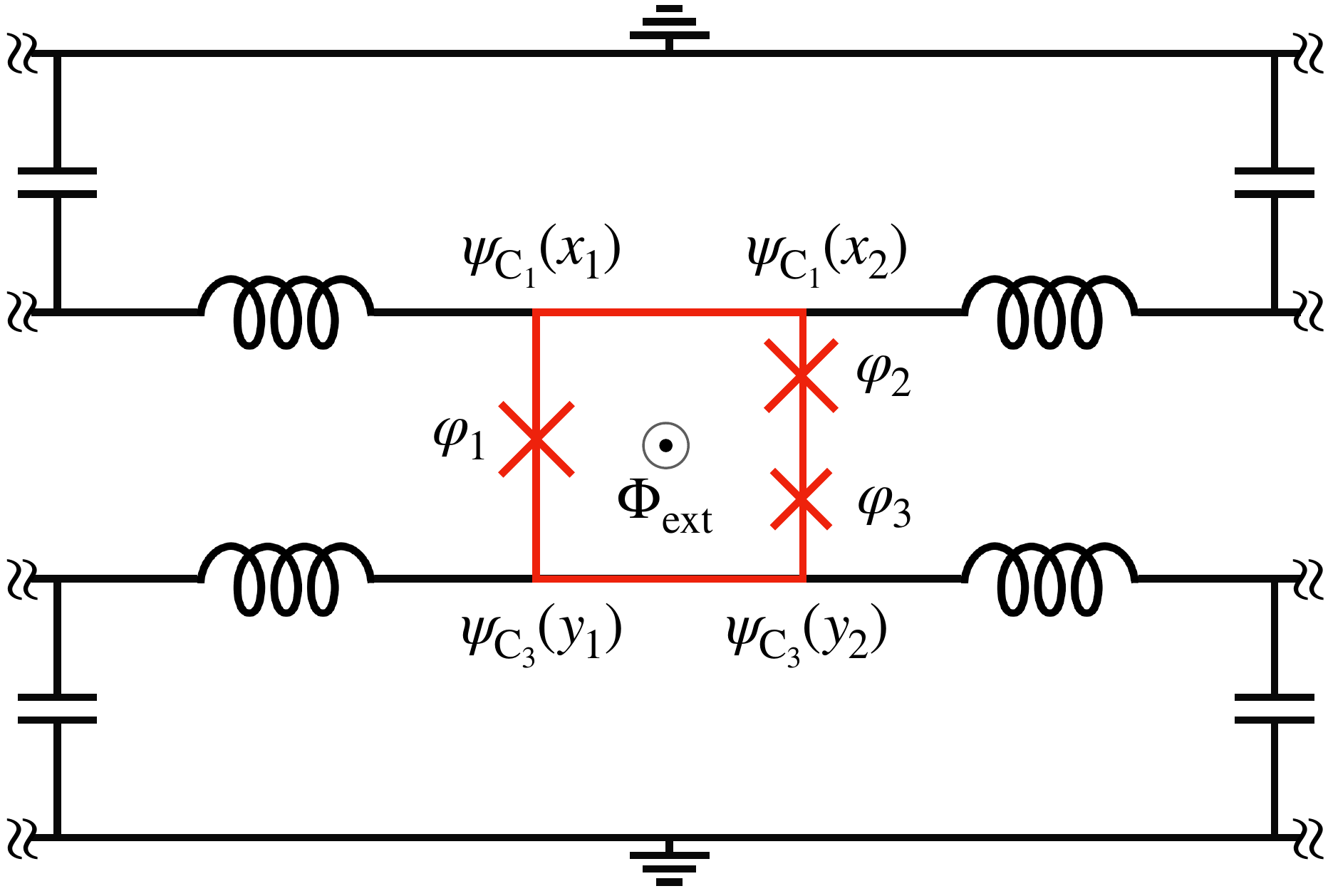}
  \caption{\textbf{Equivalent circuit diagram.} Coplanar waveguides (black lines) connected to the flux qubit (red lines).}\label{fig5}
\end{figure}
\textbf{Flux qubit-resonator}\\
The energies and electric dipole moments were calculated considering a flux qubit composed of three Josephson junctions with energies $E_{\rm J1}=E_{\rm J2}=E_{\rm J}$, and $E_{\rm J3}=\alpha E_{\rm J}$. The Hamiltonian of the flux qubit is \cite{Liu:2005eo}
\begin{equation}
  H_{\rm F}=E_{\rm C}\,P_++\frac{E_{\rm C}}{1+2\alpha}\,P_-+U(\varphi_+,\varphi_-)\,,
\end{equation}
with $U(\varphi_+,\varphi_-)=-E_{\rm J}[2\cos\varphi_+\cos\varphi_-+\alpha\cos(2\pi f+2\varphi_+)]$, having defined  $\varphi_+=(\varphi_1+\varphi_2)/2$ and $\varphi_-=(\varphi_1-\varphi_2)/2$, where $\varphi_1$ and $\varphi_2$ are the phase drops across the larger junctions. $P_+$ and $P_-$ are the conjugate momenta of $\varphi_+$ and $\varphi_-$. 
Choosing $E_{\rm J}=35\,E_{\rm C}$, $E_{\rm C}=27.1$ GHz, and $\alpha=0.8$, the dipole moment was determined by the matrix element $\langle g\vert\sin(2\pi f+2\varphi_+)\vert e\rangle$.\\

\textbf{Interaction between a flux qubit and two orthogonal coplanar waveguides}\\
The derivation of the flux qubit-resonator Hamiltonian $\hat H_{\Pi}$ is standard \cite{Bourassa:2009gy}, but here the voltage condition for the flux qubit (red loop in Fig.\,4) is $\sum_{i=1}^3\varphi_i+\Delta\psi_{\rm C_1}+\Delta\psi_{\rm C_2}=\Phi_{\rm ext}$, where $\Delta\psi_{\rm C_1}=\psi_{\rm C_1}(x_2)-\psi_{\rm C_1}(x_1)$ and $\Delta\psi_{\rm C_2}=\psi_{\rm C_2}(y_1)-\psi_{\rm C_2}(y_2)$.

 The ultrastrong coupling between a flux qubit and two superconducting coplanar stripline resonators has been experimentally realized \cite{Baust:2016ki}. However, our scheme further requires the waveguides to cross and the resonator modes not to be significantly modified by the coupling with the flux qubit. The inset in Fig.\,1a represents a sketch of the connection between the orthogonal waveguides mediated by the ancillary flux qubit. The latter is directly connected to both the center conductor of the coplanar waveguide transmission-line resonators, see also Fig.\,4.  At the insertion point, the width of the center conductor is narrower and the local inductance is larger, to enhance the coupling between the flux qubit and the resonator \cite{Bourassa:2009gy}. The three Josephson junctions forming the flux qubits must be inserted in the two tiny flux qubit arms that connect the center conductors of both waveguides. In this way, the current in the resonator flows predominantly through the center conductor constrictions of the waveguides and the resonator modes are not significantly modified. Since the distribution of the electromagnetic field is not uniform in the resonator, we suggest to fabricate waveguides with progressively narrower constrictions, in order to maintain a uniform coupling for all qubits. Alternatively, one can increase the coupling strength by inserting several Josephson junctions in the constrictions with a progressively increasing inductance along the waveguide \cite{Bourassa:2009gy}.\\

\section*{Acknowledgements}
We thank K. Inomata and R.S. Deacon for useful discussions and informations.

\end{document}